%% file: letter.tex
\documentclass[lettersize,journal]{IEEEtran}
\usepackage{amsmath,amsfonts}
\usepackage{algorithmic}
\usepackage{algorithm}
\usepackage{array}
\usepackage[caption=false,font=normalsize,labelfont=sf,textfont=sf]{subfig}
\usepackage{textcomp}
\usepackage{stfloats}
\usepackage{url}
\usepackage{verbatim}
\usepackage{graphicx}
\usepackage{cite}
\usepackage{booktabs}
\usepackage{booktabs, tabularx, makecell, array}

\begin{document}
\title{A Model-Free Detection Method for Internal Short Circuits in Single Lithium-ion Cells Using Pseudo Open-Circuit Voltage Difference}

\author{
Yangyang Xu and Chenglin Liao, \emph{Member, IEEE} \\
\textit{University of Chinese Academy of Sciences} \\
}
\maketitle
\begin{abstract}
This letter proposes a lightweight, model-free online diagnostic framework for detecting internal short circuits (ISC) in single lithium-ion cells under dynamic operating conditions. The core of the method lies in computing the first-order difference of pseudo open-circuit voltage ($\boldsymbol{\mathrm{OCV}_{\text{pseudo}}}$) to extract high-frequency deviations caused by ISC events from low-frequency polarization variations. The method relies solely on terminal voltage, current measurements, and an offline $R_0$--SOC look-up table, thereby eliminating the need for electrochemical or equivalent-circuit observers. Validated on ten real and one false fault scenarios, the proposed approach achieves a 100\% detection success rate with no missed or false alarms. In addition, the proposed method exhibits extremely low computational and memory requirements, making it highly suitable for real-time deployment in  battery management systems (BMS).
\end{abstract}


\section{Introduction}

\IEEEPARstart {W}{i}th the rapid advancement of green energy and electric transportation, lithium-ion batteries (LIBs) have become the cornerstone of modern energy storage systems~\cite{1}. However, thermal runaway (TR) remains a critical barrier to their large-scale deployment. Numerous incidents and studies have identified internal short-circuit (ISC) faults as a major trigger of TR events~\cite{2}, owing to their highly concealed and abrupt nature. Extensive research has been devoted to the early detection of both soft short circuits (SSC) and hard short circuits (HSC). In this letter, we focus on a special yet easily overlooked type of fault, transient micro internal short circuit (TMSC) , which exhibits the following two features:
\begin{itemize}
    \item A sudden voltage drop (typically 20--80\,mV) followed by rapid recovery.
    \item A very short fault duration, ranging from several seconds to tens of seconds, often exhibiting a ``fuse-off'' behavior~\cite{3}.
\end{itemize}

Such faults typically originate from lithium dendrite penetration or metallic particle intrusion through the separator, forming a temporary internal conduction path. This path may then be self-interrupted as the dendrite melts under Joule heating~\cite{4}. Although the terminal voltage quickly recovers, this process indicates underlying internal damage and should be regarded as a high-risk early warning event. If left undetected, recurrent episodes could escalate into irreversible thermal runaway incidents~\cite{5}.

A wide range of studies have sought to investigate the diagnosis of HSC faults. Hu~\textit{et al.}~\cite{6} proposed an aging-robust and disturbance-immune HSC diagnostic framework based on a model-switching scheme and recursive total least squares estimation. Zheng~\textit{et al.}~\cite{7} introduced a fast model-free method to estimate short-circuit resistance using charge loss behavior and Cramér--Rao lower bound (CRLB) theory. Qiao~\textit{et al.}~\cite{8} leveraged segment-wise dynamic time warping (DTW) distances and a GBDT classifier to detect HSC events directly from voltage data without requiring current measurements. However, most existing methods rely on module-level information, making them less applicable when only single-cell data are available. 

To address these challenges, this letter proposes a lightweight and model-free detection framework for TMSC faults in LIBs. The method constructs a pseudo open-circuit voltage ($\mathrm{OCV}_{\text{pseudo}}$) by removing the ohmic drop from terminal voltage and detects transient anomalies using its first-order difference. Compared to existing methods, the proposed approach offers the following advantages:
\begin{itemize}
    \item Applicable to single-cell measurements without dependence on module-level information or detailed modeling assumptions.
    \item Requires only basic prior data, including offline $R_0$--SOC and OCV--SOC tables, along with online voltage and current inputs.
    \item Demonstrates robust performance with accurate detection of all real faults and successful rejection of false events under dynamic conditions.
\end{itemize}

The proposed method is inspired by prior works: The concept of pseudo open-circuit voltage ($\mathrm{OCV}_{\text{pseudo}}$) is motivated by~\cite{9}.  The use of voltage differencing for anomaly detection draws inspiration from~\cite{10}.

\begin{figure}[!t]
\centering
\includegraphics[width=0.96\linewidth]{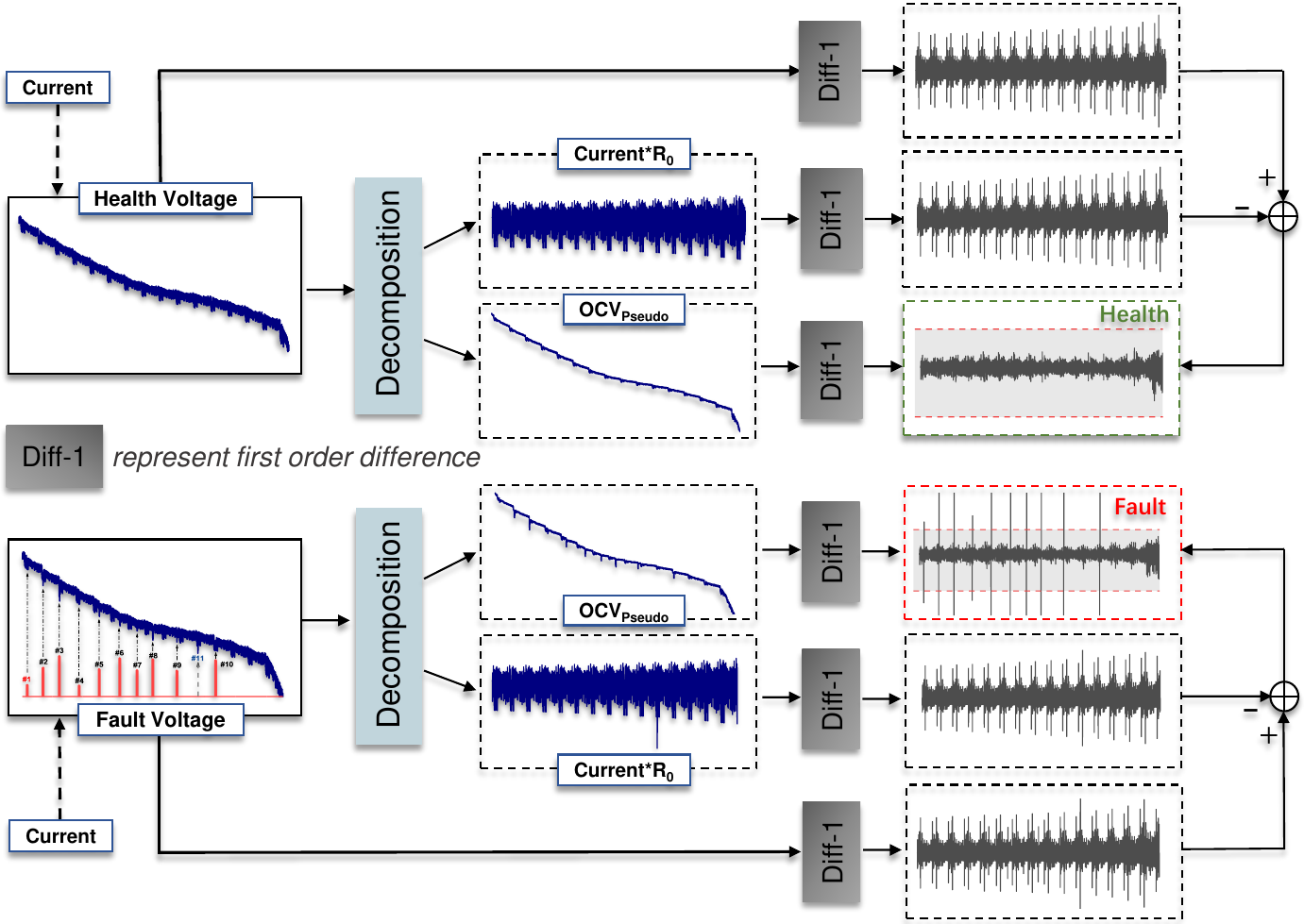}
\caption{ISC detection framework based on pseudo OCV difference.}
\label{fig2}
\end{figure}

\begin{figure}[!t]
\centering
\includegraphics[width=\linewidth]{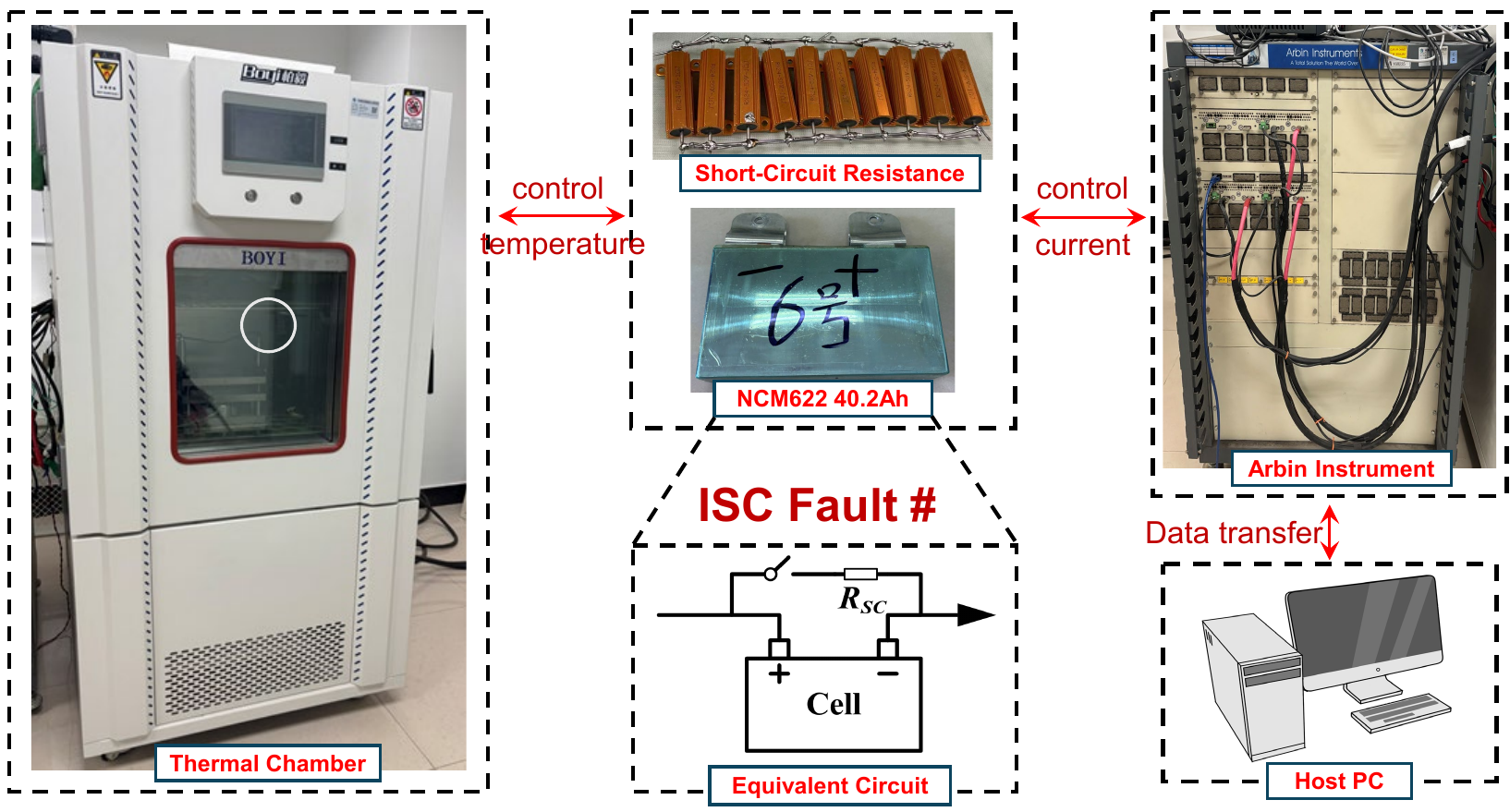}
\caption{Fault experiment equipment platform}
\label{fig1}
\end{figure}

\newpage
\section{Experiment}
This letter uses a commercial automotive battery (40.2Ah NCM622, SOH = 97.2\%) as an experimental subject.The experimental platform is illustrated in Fig. 2. We first conducted DCIR (Direct Current Internal Resistance) experiments at 5 percent SOC point intervals for offline extraction of $R_0$-SOC tables. Subsequently,the OCV-SOC table was obtained by averaging the 0.05 C charging voltage and 0.05 C discharging voltage. Fianlly, we conducted two sets of FUDS condition experiments, one for recording the voltage and current response in the healthy state, and the other for collecting the experimental data in the fault state  for comparative analysis and algorithm validation. 
\begin{figure}[!h]
\centering
\includegraphics[width=\linewidth]{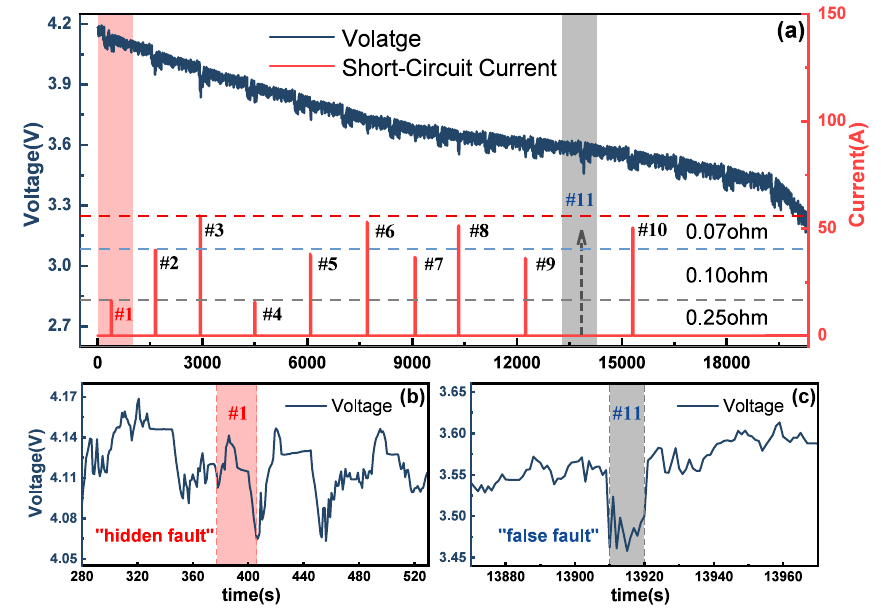}
\caption{Short-circuit fault injection timeline and corresponding voltage response during FUDS condition.}
\label{fig3}
\end{figure}

As shown in Figure 2, we used three external resistors to simulate hard short circuits inside the battery. Table 1 demonstrates the specific steps.To verify the robustness of the algorithm, we artificially designed a hidden fault and a false fault.The hidden fault was introduced by connecting a 0.25~$\Omega$ resistor during the peak charge pulse, which causes only a subtle voltage drop and lasts for a short duration, making it difficult to detect. In contrast, the false fault was generated by injecting an additional 50~A discharge pulse without any internal short circuit, simulating a pseudo-fault scenario to evaluate the algorithm's ability to avoid false positives.

\newcolumntype{C}[1]{>{\centering\arraybackslash}m{#1}}
\begin{table}[h]
  \scriptsize
  \setlength{\tabcolsep}{3pt}
  \centering
  \caption{Experiment Description}
  \label{tab:exp_desc}
  \begin{tabular}{C{1.6cm} C{3.2cm} C{1.6cm} C{1.8cm}}
    \toprule
    \textbf{Fault Type} & \textbf{Description} & \textbf{No.} & \textbf{Trigger Time} \\
    \midrule
    Severe Fault
      & \makecell[c]{\vspace{0.3mm}Triggered by an external\\0.07~$\Omega$ short-circuit resistor\vspace{0.3mm}}
      & \makecell[c]{\vspace{0.3mm}\#3, \#6, \#8, \#10\vspace{0.3mm}}
      & \makecell[c]{\vspace{0.3mm}2927s--2955s\\7708s--7738s\\10319s--10350s\\15299s--15328s\vspace{0.3mm}} \\
    \midrule
    Moderate Fault
      & \makecell[c]{\vspace{0.3mm}Triggered by an external\\0.10~$\Omega$ short-circuit resistor\vspace{0.3mm}}
      & \makecell[c]{\vspace{0.3mm}\#2, \#5, \#7, \#9\vspace{0.3mm}}
      & \makecell[c]{\vspace{0.3mm}1640s--1672s\\6083s--6113s\\9075s--9103s\\12235s--12264s\vspace{0.3mm}} \\
    \midrule
    Hidden Fault
      & \makecell[c]{\vspace{0.3mm}Triggered by a 0.25~$\Omega$ resistor\\during the peak charge pulse\vspace{0.3mm}}
      & \makecell[c]{\vspace{0.3mm}\#1, \#4\vspace{0.3mm}}
      & \makecell[c]{\vspace{0.3mm}377s--406s\\4489s--4518s\vspace{0.3mm}} \\
    \midrule
    False Fault
      & \makecell[c]{\vspace{0.3mm}Additional 50~A discharge pulse\\without external short circuit\vspace{0.3mm}}
      & \makecell[c]{\vspace{0.3mm}\#11\vspace{0.3mm}}
      & \makecell[c]{\vspace{0.3mm}13910s--13920s\vspace{0.3mm}} \\
    \bottomrule
  \end{tabular}
\end{table}

\section{Methodology}
To detect internal short-circuit (ISC) faults in lithium-ion batteries under dynamic load conditions, we propose a simple yet robust method based on pseudo open-circuit voltage ($\mathrm{OCV}_\text{pseudo}$). The approach requires no dynamic battery model and relies solely on three types of input: measured terminal voltage, current, and an offline $R_0$--SOC lookup table obtained from DCIR experiments.

At each sampling instant $k$, the pseudo OCV is calculated by subtracting the ohmic voltage drop from the measured terminal voltage:
\begin{equation}
\mathrm{OCV}_{\text{pseudo}}(k) = V_{\text{meas}}(k) + R_0(SOC(k)) \cdot I(k)
\end{equation}

This operation effectively removes the ohmic polarization, yielding a voltage estimate that closely approximates the intrinsic electrochemical state of the battery. Since the terminal voltage of the battery should stabilize around the OCV value in the absence of current excitation, the $\mathrm{OCV}_\text{pseudo}$ can be regarded as an approximation of the OCV after removing the instantaneous ohmic drop.

While the ohmic component is removed, the pseudo OCV may still contain low-frequency polarization effects such as:
\begin{itemize}
  \item \textit{Charge transfer polarization}, with time constants of several seconds;
  \item \textit{Concentration gradient (diffusion) polarization}, evolving over longer time scales.
\end{itemize}

These polarization voltages exhibit slow and smooth variations under normal conditions, with their frequency components primarily concentrated in the low-frequency range. In contrast, a hard short-circuit fault typically causes a sudden drop in terminal voltage, which is equivalent to the injection of a high-frequency step component. To extract these transient features, we compute the first-order difference of the pseudo OCV:
\begin{equation}
\Delta \mathrm{OCV}_{\text{pseudo}}(k) = \mathrm{OCV}_{\text{pseudo}}(k) - \mathrm{OCV}_{\text{pseudo}}(k-1)
\end{equation}

From a signal processing perspective, this operation is equivalent to a first-order high-pass filter: it suppresses smooth, low-frequency polarization disturbances while significantly amplifying short-term, high-frequency anomalies, making it particularly suitable for capturing transient voltage distortions caused by short-circuit faults.

We further decompose the pseudo OCV into its physical components as follows:
\begin{equation}
\mathrm{OCV}_{\text{pseudo}}(k) = \mathrm{OCV}_{\text{real}}(k) - U_{\text{polar}}(k)
\end{equation}

That is, the pseudo OCV equals the real OCV minus the low-frequency polarization drop, but still contains a residual term induced by the short-circuit current. By taking the first-order difference, we obtain:
\begin{equation}
\Delta \mathrm{OCV}_{\text{pseudo}}(k) = \Delta \mathrm{OCV}_{\text{real}}(k) - \Delta U_{\text{polar}}(k)
\end{equation}

Under healthy conditions, both $\Delta \mathrm{OCV}_{\text{real}}$ and $\Delta U_{\text{polar}}$ evolve smoothly and remain small. Under fault conditions, although $\mathrm{OCV}_{\mathrm{real}}$ and $U_{\mathrm{polar}}$ continue to vary slowly, the emergence of an internal short-circuit branch introduces an additional current $I_{\mathrm{sc}} > 0$ through the ohmic path, resulting in a new voltage drop $R_0 \cdot I_{\mathrm{sc}}$. This high-frequency voltage change is directly reflected in $\mathrm{OCV}_{\mathrm{pseudo}}$, making its first-order difference exhibit a sensitive response.
\begin{figure}[!t]
\centering
\includegraphics[width=\linewidth]{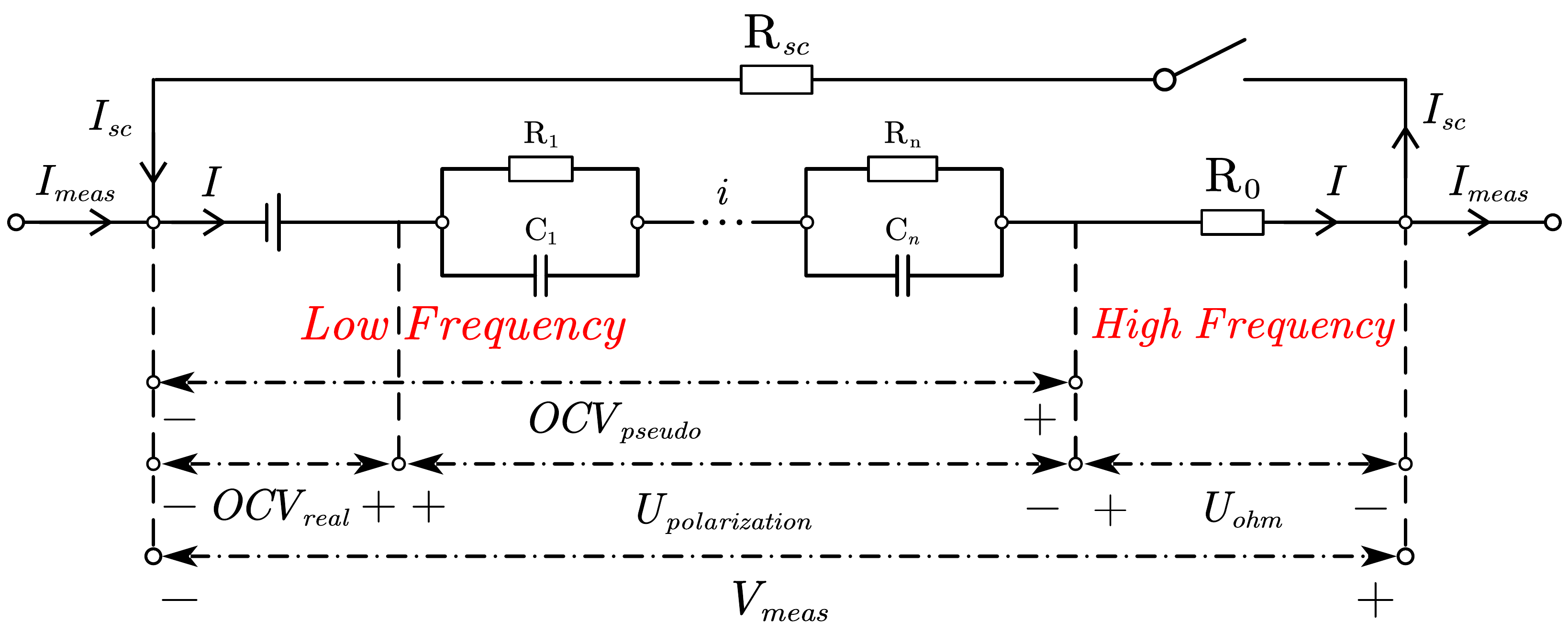}
\caption{Equivalent circuit illustrating $\mathrm{OCV}_{\text{pseudo}}$ and short-circuit current effect. The current direction is defined as discharging-positive.}
\label{fig5}
\end{figure}

This mechanism can be further understood using the equivalent circuit model illustrated in Fig.~\ref{fig5}. When the short-circuit switch closes, a low-resistance path forms, and a short-circuit current $I_{\text{sc}} > 0$ flows internally. The external output current remains $I_t$, but the actual current through $R_0$ becomes $I_t + I_{\text{sc}}$.

Consequently, the measured terminal voltage becomes:
\begin{equation}
V_{\text{meas}} = \mathrm{OCV}_{\text{real}} - U_{\text{polar}} - R_0 \cdot (I_t + I_{\text{sc}})
\end{equation}

However, the pseudo OCV is calculated assuming only the external current:
\begin{equation}
\mathrm{OCV}_{\text{pseudo}} = V_{\text{meas}} + R_0 \cdot I_t = \mathrm{OCV}_{\text{real}} - U_{\text{polar}} - R_0 \cdot I_{\text{sc}}
\end{equation}

Thus, $\mathrm{OCV}_{\text{pseudo}}$ retains a residual voltage drop caused by the unmeasurable short-circuit current. Since $I_{\text{sc}} > 0$, this term causes a noticeable drop in pseudo OCV during faults. This residual is critical: even though the fault current cannot be directly measured, its effect is embedded in the pseudo OCV, enabling reliable fault detection based on voltage anomalies alone.

To enable accurate fault timing identification, we establish a quantile-based thresholding method using healthy-condition data. Specifically, we analyze the empirical distribution of $\Delta \mathrm{OCV}_{\text{pseudo}}$ under normal operating conditions. The lower and upper thresholds are defined as:
\begin{equation}
\theta_{-} = \text{Quantile}(\Delta \mathrm{OCV}_{\text{pseudo}}, p)
\end{equation}
\begin{equation}
\theta_{+} = \text{Quantile}(\Delta \mathrm{OCV}_{\text{pseudo}}, 1 - p)
\end{equation}

where $\text{Quantile}(\cdot, p)$ returns the $p$-quantile of the input sequence, i.e., the value below which $p \times 100\%$ of the data fall. The parameter $p$ is typically chosen as 0.005, corresponding to a 99\% confidence interval.

To enhance robustness against large current pulses that may induce non-fault voltage transients, we introduce a scaling factor $\gamma > 1$ to relax the thresholds. The adjusted thresholds are defined as:
\begin{equation}
\theta_{-}' = \gamma \cdot \theta_{-}, \quad \theta_{+}' = \gamma \cdot \theta_{+}
\end{equation}

This results in a dual-sided detection rule:
\begin{itemize}
    \item If $\Delta \mathrm{OCV}_{\text{pseudo}}(k) < \theta_{-}'$, a fault onset is declared;
    \item If $\Delta \mathrm{OCV}_{\text{pseudo}}(k) > \theta_{+}'$, a fault clearance is detected.
\end{itemize}

This quantile-based approach does not assume Gaussianity and is more robust to non-symmetric fluctuations. By relaxing the threshold range via the scaling factor $\gamma$ (e.g., $\gamma = 2$), the method can effectively suppress false alarms caused by strong dynamic excitations, while preserving high sensitivity to genuine short-circuit events.

Once the fault onset time is accurately identified at sampling instant $k$ based on the pseudo OCV difference $\Delta \mathrm{OCV}_{\text{pseudo}}(k)$, we proceed to estimate the internal short-circuit resistance. Assuming that the internal short circuit causes an unobservable leakage current $I_{\mathrm{SC}}(k)$ flowing through the ohmic resistance $R_0$, the corresponding voltage drop appears as a residual in $\Delta \mathrm{OCV}_{\text{pseudo}}(k)$:
\begin{equation}
\Delta \mathrm{OCV}_{\text{pseudo}}(k) = R_0 \cdot I_{\mathrm{SC}}(k)
\end{equation}
\begin{equation}
\Rightarrow \quad I_{\mathrm{SC}}(k) = \frac{\Delta \mathrm{OCV}_{\text{pseudo}}(k)}{R_0}
\end{equation}

Considering that the short-circuit branch is connected in parallel with the battery terminals, and that the terminal voltage $V_{\mathrm{meas}}(k)$ at the fault moment reflects the voltage across the short path, the equivalent internal short-circuit resistance is given by:
\begin{equation}
R_{\mathrm{SC}}(k) = \frac{V_{\mathrm{meas}}(k)}{I_{\mathrm{SC}}(k)} = \frac{V_{\mathrm{meas}}(k) \cdot R_0}{\Delta \mathrm{OCV}_{\text{pseudo}}(k)}
\end{equation}
Due to the discrete nature of voltage sampling and the highly transient behavior of fault-induced voltage drops, it is recommended to use the first measured terminal voltage \emph{after} the fault onset as $V_{\mathrm{meas}}(k)$, ensuring that the observed drop fully reflects the effect of the internal short-circuit current.

\section{Results and Discussion}
\begin{figure}[!h]
\centering
\includegraphics[width=\linewidth]{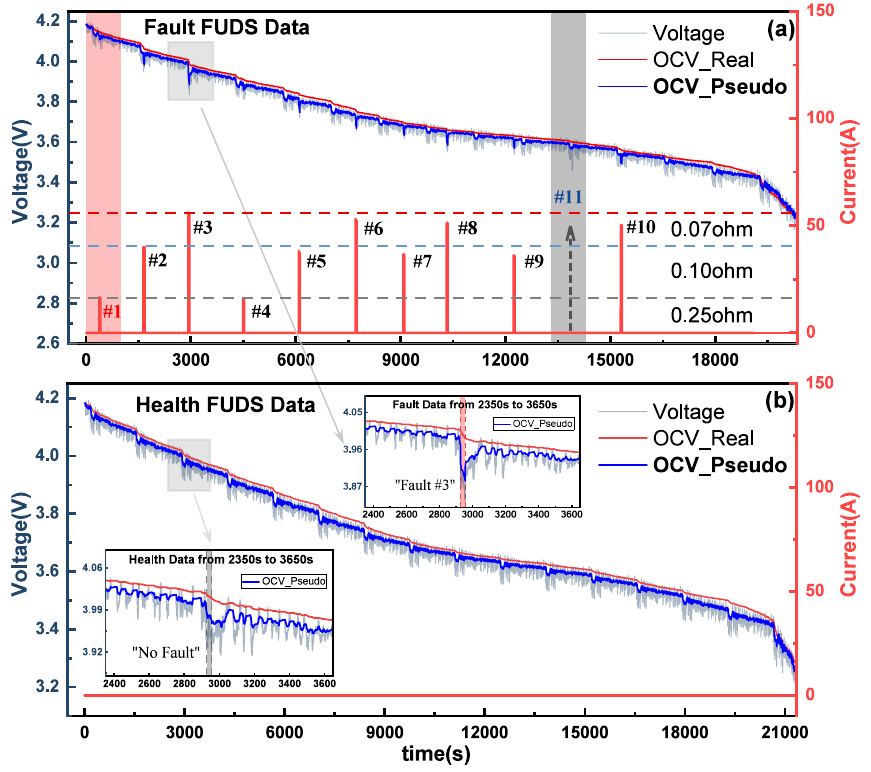}
\caption{Signatures of pseudo OCV under health and fault conditions}
\label{fig4}
\end{figure}

Fig.~\ref{fig4}(a) shows the terminal voltage response during FUDS operation with eleven fault scenarios injected. These include strong faults (e.g., \#3, \#6, \#10), hidden faults (\#1, \#4), and a false fault (\#11). When a real short circuit occurs, the pseudo OCV (blue line) drops below the real OCV (red line), and the magnitude of this deviation increases as the short-circuit resistance decreases. Notably, during the false fault \#11, although a 50~A discharge pulse is applied, the pseudo OCV remains aligned with the real OCV due to the absence of an actual internal short-circuit path, demonstrating the method's robustness against current disturbances.

Fig.~\ref{fig4}(b) provides a healthy baseline for comparison, in which the battery operates without any injected faults. In this case, the pseudo OCV closely tracks the real OCV throughout the cycle, only exhibiting minor deviations due to slow-varying polarization. The insets highlight a clear deviation during Fault \#3 and excellent alignment during a no-fault period, further verifying the stability and accuracy of the method under normal operating conditions.
\begin{figure}[!h]
\centering
\includegraphics[width=\linewidth]{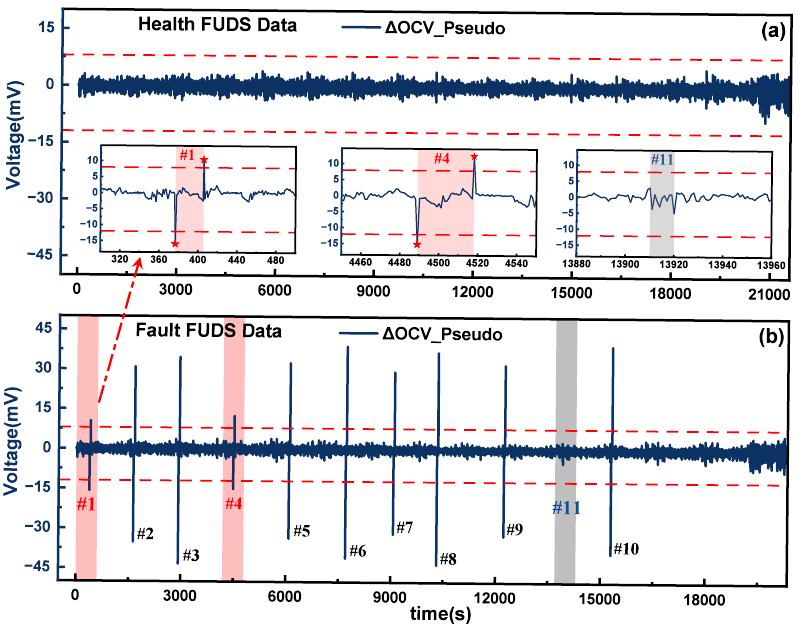}
\caption{Pseudo OCV difference under healthy and fault conditions}
\label{fig6}
\end{figure}

Fig.~\ref{fig6} presents the $\Delta \mathrm{OCV}_{\text{pseudo}}$-based detection results and quantile-based threshold construction. The upper portion of Fig.~\ref{fig6} (a) shows the first-order difference signal calculated from healthy FUDS data, which remains smooth and bounded in the absence of internal faults. Based on this healthy dataset, we compute the $p$-quantile and $(1-p)$-quantile values as statistical thresholds.To visually highlight detection performance, the lower portion of Fig.~\ref{fig6} (a) shows zoomed-in $\Delta \mathrm{OCV}_{\text{pseudo}}$ responses from three representative cases selected from Fig.~\ref{fig6} (b). Faults \#1 and \#4 represent hidden ISC events with subtle voltage drops that are still successfully detected by threshold crossings at both the fault onset and clearance. In contrast, the false fault \#11, despite involving a significant current pulse, remains entirely within the relaxed threshold range and is not misclassified.

Fig.~\ref{fig6} (b) summarizes the detection results for all eleven events. All real faults, including the hidden ones, are correctly identified by clear $\Delta \mathrm{OCV}_{\text{pseudo}}$ excursions that exceed the detection thresholds at both the fault onset and recovery moments. Meanwhile, the false fault and healthy segments remain within the threshold range, avoiding false alarms. These results confirm the method's high sensitivity to transient fault signatures and its robustness under dynamic load conditions.

\section{Conclusion}
This letter proposes a lightweight, model-free online diagnostic framework for detecting transient micro internal short circuits  in single lithium-ion cells under dynamic operating conditions. The method relies solely on terminal voltage, current measurements, and an offline $R_0$–SOC look-up table, thereby eliminating the need for electrochemical or equivalent-circuit observers. With minimal computational requirements—only one table look-up, one multiplication, and two subtractions per sampling interval—the algorithm can be readily embedded into existing MCU-based battery management systems (BMS).

\bibliographystyle{IEEEtranTIE}
\input{Letter.bbl}

\end{document}